\documentclass[twoside,final]{pro12}

\usepackage[ansinew]{inputenc}
\usepackage{amsmath}
\usepackage{graphicx} 

\usepackage{lineno}

\head{P. Zarka et al.} {Radio emission from satellite--Jupiter interactions}	

\begin{document}

\title{Radio emission from satellite--Jupiter interactions (especially Ganymede)}	
\author{P. Zarka\adress{\textsl LESIA \& USN, Observatoire Paris/CNRS/PSL, Meudon, France}$\,\,$, M.\,S. Marques$^*$\adress{\textsl INPE, S\~ao Jos\'e do Campos \& UFRN, Rio Grande do Norte, Brazil}$\,$, C. Louis$^*$, V.\,B. Ryabov\adress{\textsl Complex Systems Department, Future University Hakodate, Japan}$\,$, L. Lamy$^*$,\\ E. Echer$^\dag$, and B. Cecconi$^*$}


\maketitle

\begin{abstract}
Analyzing a database of 26 years of observations of Jupiter from the Nan\c cay Decameter Array, we study the occurrence of Io--independent emissions as a function of the orbital phase of the other Galilean satellites and Amalthea. We identify unambiguously the emissions induced by Ganymede and characterize their intervals of occurrence in CML and Ganymede phase and longitude. We also find hints of emissions induced by Europa and, surprisingly, by Amalthea. The signature of Callisto--induced emissions is more tenuous.
\end{abstract}

\section{Introduction}

Io--Jupiter decametric radio emission, that results from Alfv\'enic interaction of Io with the Jovian magnetic field, is known since 1964 [Bigg, 1964]. A similar interaction is suspected to take place with Europa whereas for Ganymede, which possesses an intrinsic magnetic field, the interaction is believed to be due to reconnection with the Jovian magnetic field [Kivelson et al., 2004]. Signatures of electron acceleration due to the above interactions have been detected in the form of ultraviolet emissions from the magnetic footprints of Io, Ganymede and Europa on Jupiter [Clarke et al., 2002]. A radio--magnetic scaling law has been proposed to quantify the energetics of satellite--planet and solar wind--magnetosphere interactions in a unified frame [Zarka, 2007; 2017]. Detecting the radio emissions induced by satellites other than Io in Jupiter's magnetic field (hereafter called simply satellite--induced radio emissions) is essential for better characterizing these interactions and test and constrain the radio--magnetic scaling law, in order to extrapolate it confidently to star--exoplanet interactions.
Early searches of satellite--induced radio emissions yielded negative results [Dulk, 1967; Kaiser and Alexander, 1973; St. Cyr, 1985]. More recent searches based on Voyager [Higgins, 2007], Galileo [Menietti et al., 1998; 2001] and Cassini [Hospodarsky et al., 2001] observations provided statistical hints of the existence of satellite--induced radio emissions, convincing in the case of Ganymede and more marginal for Europa and Callisto. The signature of Ganymede's influence was found in Galileo PWS data recorded over 2 years in the range 3.2 to 5.6~MHz. The technique used was similar to that applied by Bigg [1964] to find Io's influence on Jupiter's decametric emissions: stacked occurrence of radio emission is built in bins of a few degrees of orbital phase of the considered satellite (counted counterclockwise from the superior conjunction of the satellite as viewed from the observer -- cf. Figure~1). An influence of the satellite on the radio emissions results in a non-uniform distribution in orbital phase.

\begin{figure}[ht]
\centering
\includegraphics[width=0.65\textwidth]{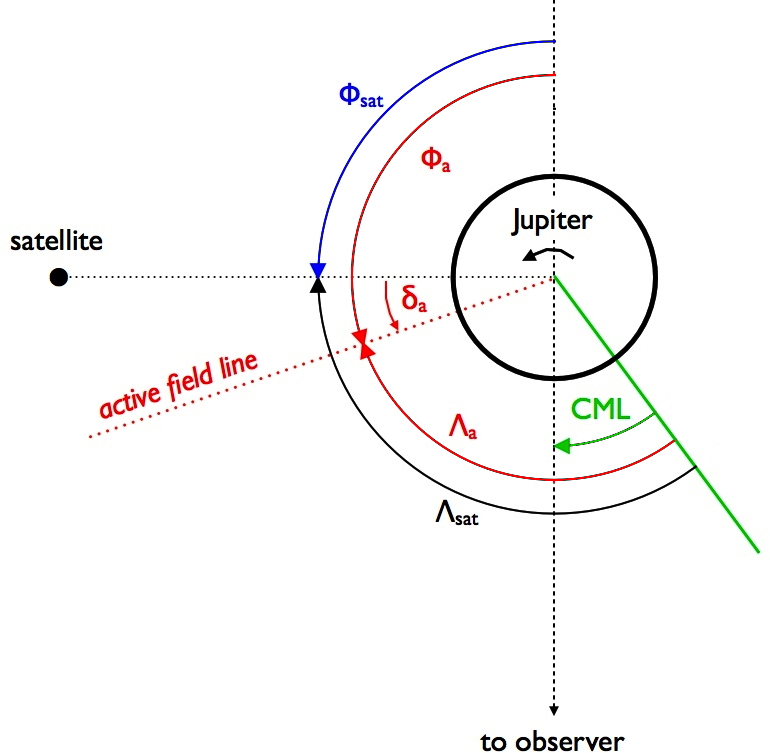}
\caption{Definition of the observer's Central Meridian Longitude (CML), and of the phase $\Phi_{sat}$ and longitude $\Lambda_{sat}$ of a Jovian satellite. Due to the propagation of the Alfv\'enic perturbation of the Jovian magnetic field by the satellite, the radio--emitting "active" field line (of phase $\Phi_{a}$ and longitude $\Lambda_{a}$) may be shifted from the instantaneous satellite field line by a lead angle $\delta a$. This lead angle is about 10$^{\circ}$ for Io due to the presence of the plasma torus [Hess et al., 2010], and it is expected to be very small for the other satellites. Adapted from Marques et al. [2017].}
\label{fig1}
\end{figure} 

\section{Io}

For Io, this non-uniform distribution takes the form of prominent peaks at two specific phases (~95$^{\circ}$ and 240$^{\circ}$ -- Figure~2b). The recent construction and analysis of a database of Jupiter's decametric emissions detected over 26 years of observations with the Nan\c cay Decameter Array (NDA) in the 10--40 MHz range [Boischot et al., 1980; Lamy et al., 2017] resulted in a fine and homogeneous characterization of Io--induced and Io--independent emissions [Marques et al., 2017]. It was found that Io--induced emissions are detected ~5.9\% of the time, versus 4.6\% for Io--independent emissions. The occurrence of Io--induced emissions is so high that these emissions show up very clearly in diagrams where occurrence probability -- as defined in [Marques et al., 2017] -- is plotted versus two coordinates (cf. Figure~2a): $\Phi_{Io}$ (the above described Io phase) and the CML (Central Meridian Longitude = observer's Jovicentric longitude). Restricted regions of (CML, $\Phi_{Io}$) appear, that have been labeled A, B, C, D and interpreted as the radio sources at both (Northern and Southern) footprints of Io's magnetic flux tube emitting along hollow conical sheets seen by a distant observer from the West (B, D) or East (A, C) limb of Jupiter, as illustrated in Figure~3 (actually in the case of Io, the radio--emitting or --active flux tube is slightly shifted from the instantaneous Io flux tube -- cf. Figure~1). This interpretation derives from the fact that, when plotted versus Io's longitude $\Lambda_{Io}$ (defined in Figure~1), occurrence from all "Io" regions of the (CML, $\Phi_{Io}$) plane gather in a single broad region in $\Lambda_{Io}$ (Figure~2c). The hemisphere where the source lies can be identified mainly from the sense of circular polarization of the emission (Right-Hand from the North, Left-Hand from the South), not discussed further here. Sub-regions labeled A', A'', and B' are discussed in Marques et al. [2017] and are beyond the scope of this paper.

\begin{figure}[ht]
\centering
\includegraphics[width=0.9\textwidth]{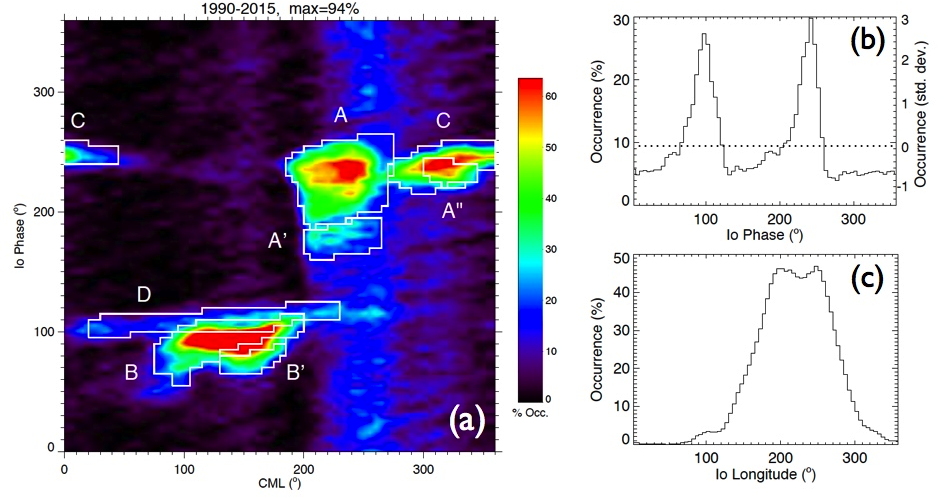}
\caption{(a) Occurrence probability of Jupiter's decametric radio emissions detected over 1990--2015 with the NDA, as a function of CML and $\Phi_{Io}$, in $5^{\circ} \times 5^{\circ}$ bins. Io--induced emissions concentrate in the white boxes (where their occurrence probability is $\geq 10$\%). Io--independent emissions are the broad bands dependent on the CML only. (b) Occurrence probability versus $\Phi_{Io}$ only, resulting from the integration of (a) over the CML. The right-hand scale is in units of its standard deviation relative to its mean (dotted line). (c) Occurrence probability of Io--induced emissions only as a function of $\Lambda_{Io}=$CML$+180^{\circ}-\Phi_{Io}$.}
\label{fig2}
\end{figure} 

Occurrence reaches 94\% (i.e. nearly permanent emission) at the center of the Io-B region. The white boxes superimposed on Figure~2a indicate the geometries of observation for which the occurrence probability of Io--induced emissions is $\geq 10$\% (in $5^{\circ} \times 5^{\circ}$ bins). These boxes can be used to plan Jupiter observations, and their coordinates can be found on the NDA web site (https://www.obs-nancay.fr/-Le-reseau-decametrique-.html). Broad bands of emission occurrence restricted in CML but independent of $\Phi_{Io}$ can be distinguished in Figure~2a. When polarization is taken into account, four such bands -- partly overlapping in CML -- can be identified (see Table~5 of Marques et al. [2017]), that correspond to decametric radio emissions independent of Io (i.e. of auroral origin or induced by other satellites), labeled non-Io-A to non-Io-D depending on their circular polarization and their Jovian limb of origin as revealed by the curvature of the emission in the time--frequency plane (see Figure~3 and details in Marques et al. [2017]).

\begin{figure}[ht]
\centering
\includegraphics[width=0.9\textwidth]{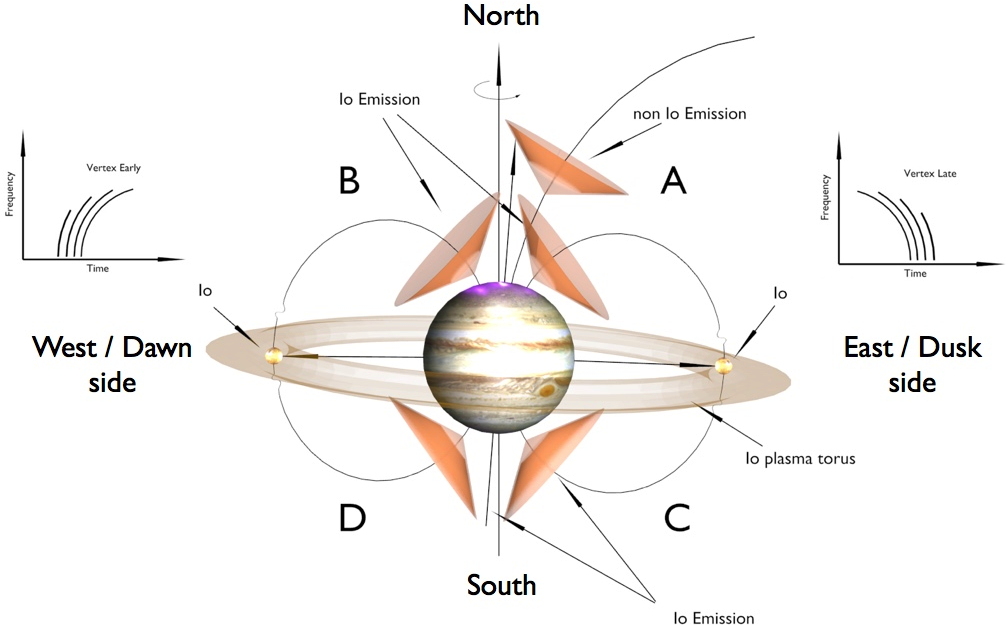}
\caption{Geometry and nomenclature (A...D) of Io and non-Io emissions, and corresponding arc shapes in the t--f plane. Emission is produced along the displayed conical sheets, and thus observable only when the source is near a limb of Jupiter. Dawnside (West) sources produce vertex--early arcs whereas duskside (East) sources produce vertex--late arcs, as explained in Hess et al. [2014]. The magnetic field line connected to Io is sketched as the active, radio--emitting field line but actually the active field line leads the instantaneous Io field line by several degrees (cf. Fig.~1). Adapted from Marques et al. [2017].}
\label{fig3}
\end{figure} 

\section{Ganymede}

Although statistically significant, the influence of Ganymede on Jupiter's radio emissions as detected from Galileo observations is weak and noisy. It shows up as a nearly sinusoidal modulation of occurrence as a function of $\Phi_{Ga}$ (the orbital phase of Ganymede), but is very difficult to identify as regions of the (CML, $\Phi_{Ga}$) diagram [Menietti et al., 1998; Hospodarsky et al., 2001].

We have searched the radio emission induced by Ganymede in the 26 years NDA database with the same technique. The database has the advantage that each emission event is identified with its type (Io-A, -B, -C, ... or non-Io-A, -B, -C, -D), allowing us to efficiently exclude all Io--induced emissions in order to search for the Ganymede influence in non-Io emission events. This is necessary because of the 1:4 orbital resonance of Ganymede with Io (cf. Table \ref{Tab.1}). The NDA database also covers a much longer time-interval than the Galileo data. The distribution of emission occurrence probability versus the CML and $\Phi_{Ga}$ is displayed in Figure~4a. Regions of enhanced occurrence show up, but the diagram is dominated by the high occurrence that covers the CML range $190^{\circ}-285^{\circ}$. This emission is non-Io-A emission independent of $\Phi_{Ga}$, that corresponds thus to truly auroral emission from Jupiter's northern hemisphere / dusk side (aurora is known to be very active in this sector -- see e.g. Hess et al. [2014]). In 4b we have reproduced Figure~4a except that the occurrence in the CML range $190^{\circ}-285^{\circ}$ has been divided by 10. Regions of enhanced occurrence in (CML, $\Phi_{Ga}$) show up much more clearly, and analysis by individual non-Io component (i.e. non-Io-A to non-Io-D) allowed us to define the white boxes superimposed on Figures~4a and 4b. These boxes are labeled from the non-Io component in which the corresponding emissions show up (e.g. Ganymede-A emissions correspond to the peak of occurrence of non-Io-A emission events observed as a function of $\Phi_{Ga}$).

\begin{table}[h]
\caption{Orbital parameters, resonances, and magnetic flux tube footprint latitudes (in a dipolar field) of Io, Europa, Ganymede, Callisto and Amalthea.}\label{Tab.1}
\begin{center}
\begin{tabular}{|c|c|c|c|c|} 
\hline 
Satellite &	Orbital  &	Orbital  & Resonance  & Footprint's magnetic  \\
  &	 distance (R$_J$) &	 period (h) &  w.r.t. Io & latitude ($^\circ$) \\
\hline	
Io			& 5.90	& 42.46	& 1.00	& 66. \\
Europa		& 9.39	& 85.22	& 2.01	& 71. \\
Ganymede	& 14.97	& 171.72	& 4.04	& 75. \\
Callisto		& 26.33	& 400.56	& 9.43	& 79. \\
Amalthea		& 2.54	& 11.95	& 0.28	& 51. \\
\hline
\end{tabular}
\end{center}
\end{table}

\begin{figure}[ht]
\centering
\includegraphics[width=0.9\textwidth]{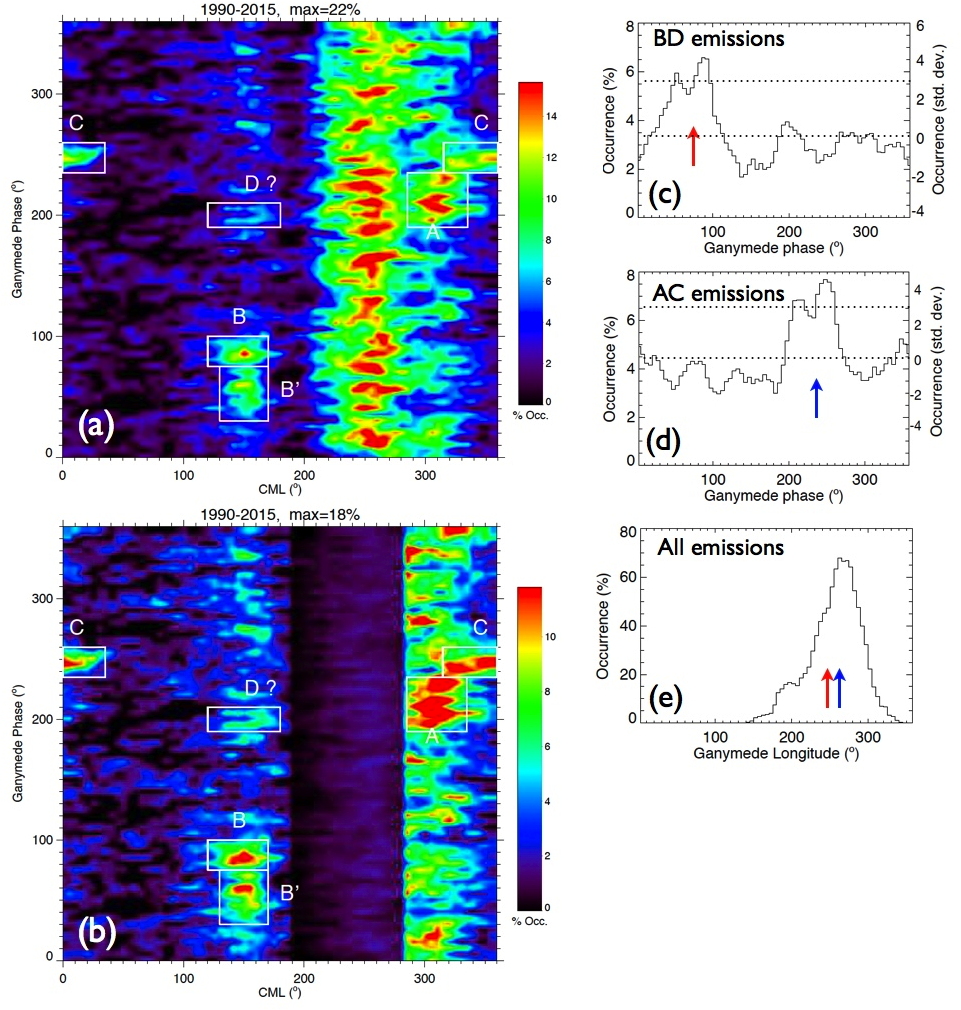}
\caption{(a) Occurrence probability of Io--independent decametric radio emissions detected over 1990--2015 with the NDA, as a function of CML and $\Phi_{Ga}$ (the orbital phase of Ganymede), in $5^{\circ} \times 5^{\circ}$ bins. (b) Same as (a) except that the values in the CML range $190^{\circ}-285^{\circ}$ have been divided by 10. Ganymede--induced emissions show up as concentrations in the white boxes, also drawn in (a), that were defined by plotting the occurrence probability separately for non-Io-A, -B, -C, and -D emissions. Ganymede-D emissions are marginally identified. (c) Occurrence probability versus $\Phi_{Ga}$ in the CML range $120^{\circ}-180^{\circ}$, smoothed over a 5 bins boxcar window. (d) Occurrence probability versus $\Phi_{Ga}$ in the CML range $290^{\circ}-35^{\circ}$ (i.e. $290^{\circ}-395^{\circ}$), smoothed over 5 bins. The right-hand scale of (c,d) is in units of standard deviations, with the dotted line indicating the mean and mean$+3 \sigma$ levels. (e) Occurrence probability of Ganymede--induced emissions only as a function of $\Lambda_{Ga}=$CML$+180^{\circ}-\Phi_{Ga}$. The peaks in (c) and (d) identified as colored arrows merge in $\Lambda_{Ga}$.}
\label{fig4}
\end{figure} 

Figure~4c shows the dependence versus $\Phi_{Ga}$ of the emission events occurring in the CML range that encompasses the B, B' and D boxes, i.e. $120^{\circ}-180^{\circ}$. Figure~4d shows the dependence versus $\Phi_{Ga}$ of the emission events occurring in the CML range that encompasses the A and C boxes, i.e. $290^{\circ}-35^{\circ}$ (or $290^{\circ}-395^{\circ}$). In both cases a prominent peak is observed, at a $>4 \sigma$ significance level, that is further enhanced by the broad width of the peak at a level $>3 \sigma$. The dual--peak shapes come from the fact that these curves actually mix emissions from the same limb but from different hemispheres (i.e. B+B' emissions on Figure~4c -- D emissions contribute for a smaller peak at $\Phi_{Ga}\sim200^\circ$ --, and A+C emissions on Figure~4d). The positions of the peaks in $\Phi_{Ga}$ is close to -- but actually slightly lower than -- that of Io--induced emission peaks in $\Phi_{Io}$. This could be due to the combination of the topology of Jovian field lines interacting with Ganymede (at 15~R$_J$ radial distance) with a quasi--null lead angle due to the absence of a plasma torus around Ganymede's orbit. Confirmation of the physical origin of these peaks comes from the fact that, when the contents of the white boxes of Figures~4a,b is plotted versus the longitude of Ganymede $\Lambda_{Ga}$, they gather in a single broad region in $\Lambda_{Ga}$ (Figure~4e) that coincides with the higher half of the distributions of Io emissions versus $\Lambda_{Io}$ (Figure~2c).
The peaks in Figures~4c and 4d match the maxima found in Galileo data by Menietti et al. [1998] and Hospodarsky et al. [2001], but they have a much better signal--to--noise ratio and characterize in greater details the occurrence of the emission, as further illustrated by Figures~4a,b. Also, emission events attributed to the Ganymede--Jupiter interaction are individually identified, allowing us to study their dynamic spectrum, polarization, hemisphere and limb of origin, and compare them to magnetic field models [e.g. Hess et al., 2011, 2017a] and ExPRES simulations [Hess et al., 2008, 2017b; Louis et al., 2017a,b]. This will be the subject of a further study.

The boundaries of the Ganymede occurrence boxes in Figures~4a,b are listed in Table \ref{Tab.2}. They are consistent with those found by Louis et al. [2017a,b] by comparing ExPRES simulations with observations from Voyager and Cassini (+NDA and Wind), the differences being attributed to the different frequency ranges of the observations studied. Table \ref{Tab.2} also lists the number of emission events detected in the 26 years NDA database for each Ganymede--induced emission type. This number is approximately 1/10th of the number of Io--induced and Io--independent emission events. The energetics of the Ganymede--Jupiter interaction, compared to that of the Io--Jupiter interaction, is studied in details in Zarka et al. [2017]. Maximum occurrence probability of Ganymede--induced emissions is above 10\% in restricted regions of the (CML, $\Phi_{Ga}$) plane and reaches 18\% in the Ganymede-B region.

\begin{table}[h]
\caption{Boundaries of Ganymede--induced emission regions in the (CML, $\Phi_{Ga}$) plane and number of emissions within each box.}\label{Tab.2}
\begin{center}
\begin{tabular}{|l|l|l|l|} 
\hline 
Emission type &	CML range &	$\Phi_{Ga}$ range & Number of emissions\\
\hline	
Ganymede-A &  $285^{\circ}-335^{\circ}$ &		$190^{\circ}-235^{\circ}$	&	125\\
Ganymede-B &	$120^{\circ}-170^{\circ}$ &		$75^{\circ}-100^{\circ}$ & 61\\
Ganymede-B' & $130^{\circ}-170^{\circ}$ &		$30^{\circ}-75^{\circ}$ &	91\\
Ganymede-C &  $315^{\circ}-395^{\circ}$ &		$235^{\circ}-260^{\circ}$ & 63\\
Ganymede-D? &	$120^{\circ}-180^{\circ} (-210^{\circ})$ & $190^{\circ}-210^{\circ}$ & 7 (+10)\\
\hline
 & & & Total 347 (+10)\\
\hline
\end{tabular}
\end{center}
\end{table}

\section{Europa}

The same technique has been used for looking for Europa--induced emission, again excluding Io--induced emissions due to the 1:2 orbital resonance of Europa with Io (cf. Table \ref{Tab.1}). We have selected the emissions independent of both Io and Ganymede, detected over 1990--2015 with the NDA. Figure~5a displays the occurrence probability of the part of these emissions originating from the northern hemisphere (i.e. A and B, Right-Hand polarized) as a function of CML and $\Phi_{Eu}$. Figures~5b displays the occurrence probability of southern emissions (i.e. C and D, Left-Hand polarized). Non-uniform distributions show up as a function of $\Phi_{Eu}$ in some CML ranges.

\begin{figure}[ht]
\centering
\includegraphics[width=0.9\textwidth]{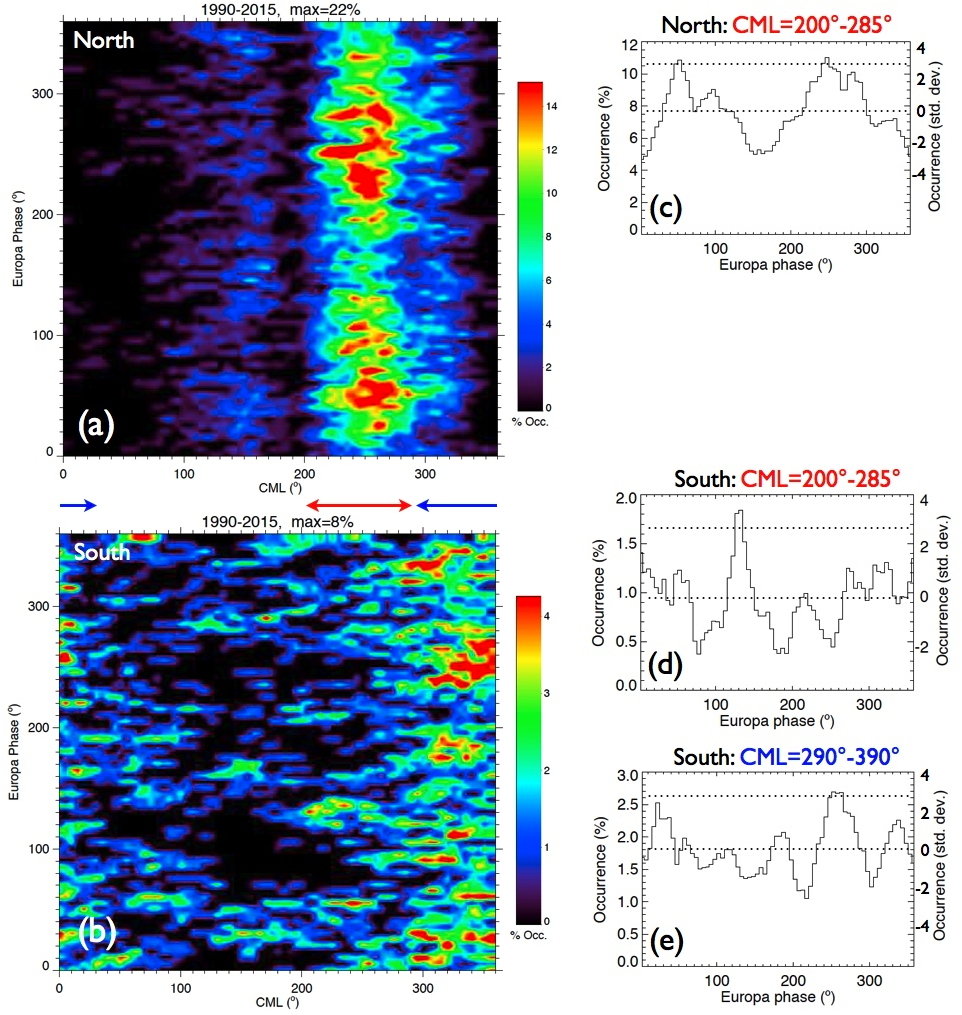}
\caption{(a) Occurrence probability of Jupiter's northern decametric radio emissions independent of Io and Ganymede, detected over 1990--2015 with the NDA, as a function of CML and $\Phi_{Eu}$ (the orbital phase of Europa), in $5^{\circ} \times 5^{\circ}$ bins. (b) Same as (a) for southern emissions. (c) Occurrence probability of northern emissions versus $\Phi_{Eu}$ in the CML range $200^{\circ}-285^{\circ}$ (indicated by the red arrow between (a) and (b)), smoothed over 5 bins. (d) Same as (c) for southern emissions. (e) Occurrence probability of southern emissions versus $\Phi_{Eu}$ in the CML range $290^{\circ}-390^{\circ}$ (indicated by the blue arrows), smoothed over 5 bins. The right-hand scale of (c,d,e) is in units of standard deviations, with the dotted line indicating the mean and mean$+3 \sigma$ levels.}
\label{fig5}
\end{figure} 
 
Figure~5c displays the dependence versus $\Phi_{Eu}$ of the northern emission events occurring in the CML range $200^{\circ}-285^{\circ}$ (smoothed over 5 bins). A strong sinusoidal modulation is visible, not inconsistent with but much more prominent than that found by Higgins [2007]. Peaks at $>3 \sigma$ level are visible on top of the sinusoidal modulation, at $\Phi_{Eu} \sim 50^{\circ}$ and 250$^{\circ}$. Note however that this modulation represents a $\pm 2.5$\% modulation superimposed on a constant background at $\sim 5$\% (that corresponds to the non-Io-A auroral radio emission from Jupiter). The southern counterpart of this modulation (Figure~5d) shows a well-defined peak $>3 \sigma$ at $\Phi_{Eu}=130^{\circ}$, that has a $\sim 1$\% amplitude on top of a constant background at $\sim 0.7$\%. In the CML range $290^{\circ}-390^{\circ}$ where most of the southern emissions take place, a broad peak $>3 \sigma$ is detected around $\Phi_{Eu}=260^{\circ}$.

Could these modulations and peaks be due to a residual dependence of non-Io emissions versus $\Phi_{Io}$ combined with the Europa--Io resonance ? Figure 7b of Marques et al. [2017] shows that non-Io emissions are very little polluted by residual Io--induced emissions, but conversely that a few \% of non-Io emissions have been catalogued as Io--induced in the database, resulting in a broad gap versus $\Phi_{Io}$. This gap is more pronounced in the CML range $\ge290^\circ$, where 0 (in Figure 5a) or 1 (in Figures 5b,e) peak appears. It is thus unlikely that in the CML range $200^\circ - 285^\circ$ the smaller gap in non-Io emission occurrence causes the prominent peaks observed here for Europa.

Overall, this behaviour demonstrates the existence of Europa--induced decametric radio emissions, also found in spacecraft data by Louis et al. [2017a,b].

\section{Callisto}

The same analysis as for Europa is carried as a function of the orbital phase of Callisto $\Phi_{Ca}$ in Figure~6. A peak $>3 \sigma$ shows up in Figure~6c, but it has a width of one 5$^{\circ}$ bin only, thus it may result from random fluctuations of the distribution of occurrence probability in Figure~6a. No other hint of Callisto--induced decametric emission is observed in the northern hemisphere. In the southern hemisphere, although fluctuations of the occurrence probability versus $\Phi_{Ca}$ are less pronounced than versus the Europa phase, a roughly sinusoidal modulation is observed in the CML range $290^{\circ}-390^{\circ}$ (Figure~6e) where most of the southern emissions take place, as well as in the CML range $80^{\circ}-180^{\circ}$ (Figure~6d). Actually, the structure of the peaks in Figure~6e is very similar to that observed by Galileo, as shown in Fig.~5 of Hospodarsky et al. [2001]. Our results thus support the likeliness of a weak control of Jupiter's decametric radio emission by Callisto without bringing a definitive confirmation.

We believe that this result is not due to the non--resonance of Callisto with Io (Table \ref{Tab.1}) (which would implicitly suggest that the detections of the Ganymede and Europa effects are spurious), but to the very weak strength of the Callisto interaction with Jupiter's magnetic field [Zarka, 2007].

\begin{figure}[ht]
\centering
\includegraphics[width=0.9\textwidth]{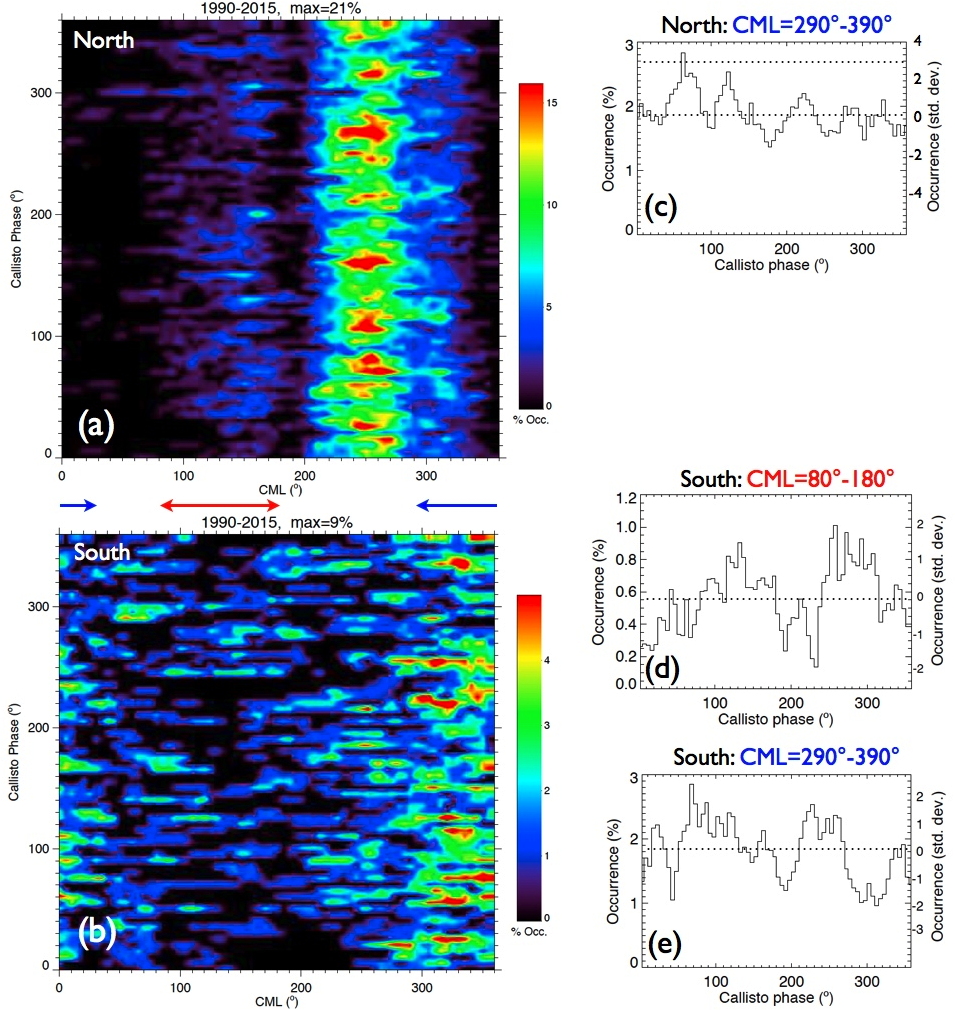}
\caption{(a) Occurrence probability of Jupiter's northern decametric radio emissions independent of Io and Ganymede, detected over 1990--2015 with the NDA, as a function of CML and $\Phi_{Ca}$ (the orbital phase of Callisto), in $5^{\circ} \times 5^{\circ}$ bins. (b) Same as (a) for southern emissions. (c) Occurrence probability of northern emissions versus $\Phi_{Ca}$ in the CML range $290^{\circ}-390^{\circ}$ (indicated by the blue arrows between (a) and (b)), smoothed over 5 bins. (d) Occurrence probability of southern emissions versus $\Phi_{Ca}$ in the CML range $80^{\circ}-180^{\circ}$ (indicated by the red arrow), smoothed over 5 bins. (e) Same as (c) for southern emissions. The right-hand scale of (c,d,e) is in units of standard deviations, with the dotted line indicating the mean and mean$+3 \sigma$ levels.}
\label{fig6}
\end{figure} 

\section{Amalthea}

Finally, Figure~7a to 7g display the results of the same analysis carried as a function of the orbital phase of Amalthea $\Phi_{Am}$. In several CML ranges in both hemispheres, a small fraction of the emissions is modulated by $\Phi_{Am}$ with a statistically significant amplitude. The modulation is more prominent in the southern hemisphere because it is superimposed on a smaller background of Jovian auroral radio emissions, but it is also visible in the northern hemisphere at a level $>3 \sigma$.

\begin{figure}[ht]
\centering
\includegraphics[width=0.9\textwidth]{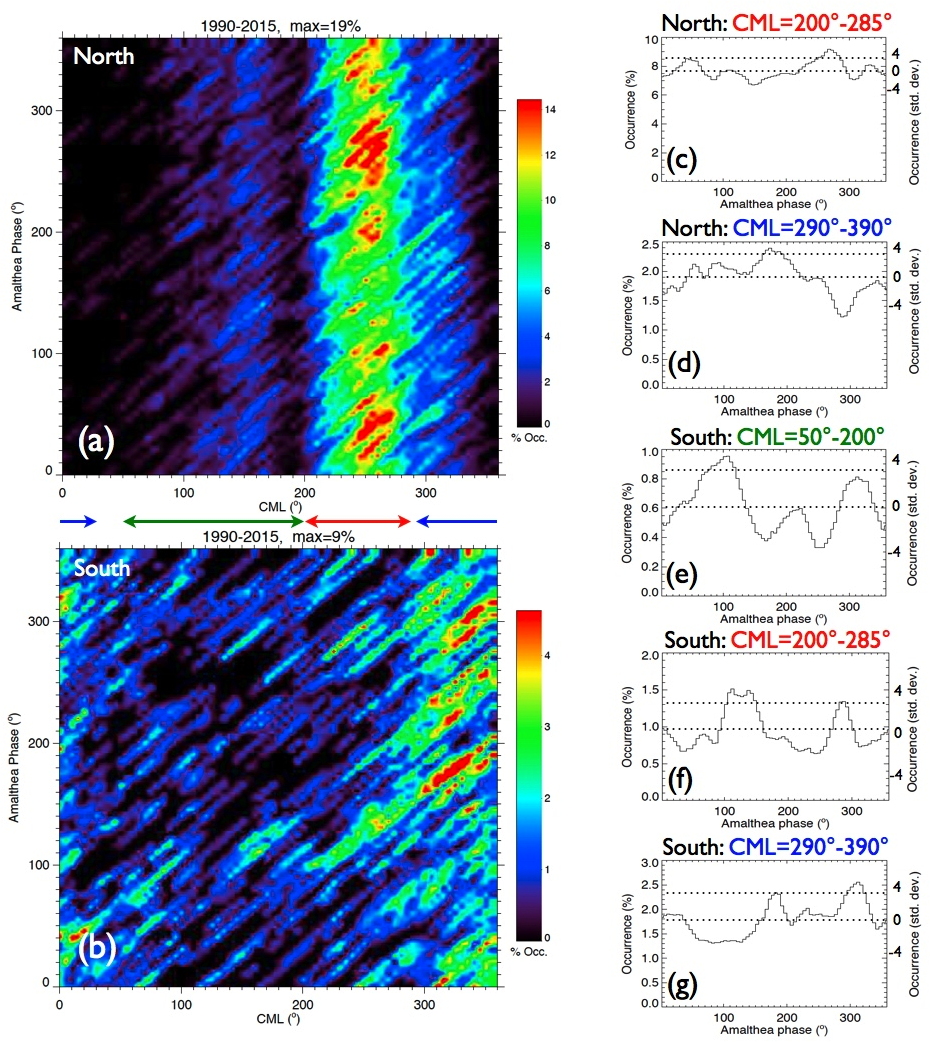}
\caption{(a) Occurrence probability of Jupiter's northern decametric radio emissions independent of Io and Ganymede, detected over 1990--2015 with the NDA, as a function of CML and $\Phi_{Am}$ (the orbital phase of Amalthea), in $5^{\circ} \times 5^{\circ}$ bins. The slanted lines (of slope $\sim$0.83) result from the combination of Amalthea's orbital period $\sim$12 h with Jupiter's rotation $\sim$10 h. (b) Same as (a) for southern emissions. (c) Occurrence probability of northern emissions versus $\Phi_{Am}$ in the CML range $200^{\circ}-285^{\circ}$ (indicated by the red arrow between (a) and (b)), smoothed over 5 bins. (d) Occurrence probability of northern emissions versus $\Phi_{Am}$ in the CML range $290^{\circ}-390^{\circ}$ (indicated by the blue arrows), smoothed over 5 bins. (e) Occurrence probability of southern emissions versus $\Phi_{Am}$ in the CML range $50^{\circ}-200^{\circ}$ (indicated by the green arrow), smoothed over 5 bins. (f) Same as (c) for southern emissions. (g) Same as (d) for southern emissions. The right-hand scale of (c...g) is in units of standard deviations, with the dotted line indicating the mean and mean$+3 \sigma$ levels.}
\label{fig7}
\end{figure} 

Amalthea is a small non-spherical Jovian moon of average diameter 167~km, not part of the Galilean satellites family. As it orbits at only 2.54 Jovian radii from Jupiter and is thus connected to latitudes $\sim 51^{\circ}$ at the surface of the planet (cf. Table \ref{Tab.1}), it is not a-priori expected to modulate the decametric radio emissions thought to originate from much higher latitudes [Zarka, 1998]. Nevertheless, Dulk [1967] looked for its influence (without finding any) whereas Arkhypov and Rucker [2007] found that part of the so-called "modulation lanes" of Jupiter's decametric emission can be explained by plasma inhomogeneities at the orbit of Amalthea. Another hint comes from Fillius et al. [1975], who found absorption signatures in Pioneer 11 observations of Jupiter's radiation belts that could be attributed to a ring or torus of particles at the orbit of Amalthea.

It should be noted that the orbital period of Amalthea is 0.4982 mean solar day, whereas long-term ground-based observations of Jupiter -- such as the NDA ones -- are carried out daily with a periodicity equal to the synodic period between a sidereal day and the orbital period of Jupiter, i.e. $0.9974 = 2 \times 0.4987$ mean solar day. The relative difference between Amalthea's orbital period and the half-period of daily observations is thus $10^{-3}$, so that it may be difficult to separate the two from observations spread over less than 1000 days. However, the 26 years NDA database extends over $>9000$ days, so that it is possible to separate these two periods (we have checked that two well separated peaks at 0.4982 and 0.4987 actually exist in the long-term harmonic analysis of the recorded emissions), so that the results from Figure~7 may reveal a real influence of Amalthea on Jovian decametric radio emissions. The position of the various peaks versus $\Phi_{Am}$ deserves a further study that may eventually constrain the physical nature of this influence.

\section{Conclusions and perspectives}

The analysis of the 26 years NDA database from Marques et al. [2017] allowed us to (i) produce a reference map of the occurrence probability of Io--induced and Io--independent radio emissions (Figure~2a); (ii) detect unambiguously and characterize the occurrence probability of Ganymede--induced radio emissions versus CML, Ganymede phase $\Phi_{Ga}$ and longitude $\Lambda_{Ga}$; (iii) demonstrate the existence of Europa--induced decametric radio emissions; (iv) support the likeliness of a weak control of Jupiter's decametric radio emission by Callisto; (v) strongly suggest the existence of an influence of Amalthea on Jovian decametric radio emissions. The detailed nature and energetics of these interactions and the possible role of orbital resonances will be the subject of subsequent studies.

\section*{References}
\everypar={\hangindent=1truecm \hangafter=1}

Arkhypov, O.\,V., and H.\,O. Rucker, Amalthea's modulation of Jovian decametric radio emission, \textsl{Astron. Astrophys.}, \textbf{467}, 353--358, 2007.

Bigg, E.\,K., Influence of the satellite Io on Jupiter's decametric emission, \textsl{Nature}, \textbf{203}, 1008--1010, 1964.

Boischot, A., C. Rosolen, M.\,G. Aubier, G. Daigne, F. Genova, Y. Leblanc, A. Lecacheux, J. de la No\"e, and B.\, M. Pedersen, A new high gain, broadband steerable array to study Jovian decametric emission, \textsl{Icarus}, \textbf{43}, 399--407, 1980.

Clarke, J.\,T., J. Ajello, G. Ballester, L. Ben Jaffel, J. Connerney, J.--C. G\'erard, G.\,R. Gladstone, D. Grodent, W. Pryor, J. Trauger, and J.\,H. Waite Jr, Ultraviolet emissions from the magnetic footprints of Io, Ganymede and Europa on Jupiter, \textsl{Nature}, \textbf{415}, 997--1000, 2002.

Dulk, G.\,A., Lack of effects of satellites Europa, Ganymede, Callisto, and Amalthea on the decametric radio emission of Jupiter, \textsl{Astrophys. J.}, \textbf{148}, 239--248, 1967.

Fillius, R.\,W, C.\,E. McIlwain, and A. Mogro--Campero, Radiation belts of Jupiter: A second look, \textsl{Science}, \textbf{188}, 4187, 465--467, 1975.

Hess, S.\,L.\,G., B. Cecconi, and P. Zarka, Modeling of Io--Jupiter decameter arcs, emission beaming and energy source, \textsl{Geophys. Res. Lett.}, \textbf{35}, L13107, 2008.

Hess, S.\,L.\,G., A. P\'etin, P. Zarka, B. Bonfond, and B. Cecconi, Lead angles and emitting electron energies of Io--controlled decameter radio arcs, \textsl{Planet. Space Sci.}, \textbf{58}, 1188--1198, 2010.

Hess, S.\,L.\,G., B. Bonfond, P. Zarka, and D. Grodent, Model of the Jovian magnetic field topology constrained by the Io auroral emissions, \textsl{J. Geophys. Res.}, \textbf{116}, A05217, 2011.

Hess, S.\,L.\,G., E. Echer, P. Zarka, L. Lamy, and P. A Delamere, Multi-instrument study of the Jovian radio emissions triggered by solar wind shocks and inferred magnetospheric subcorotation rates, \textsl{Planet. Space Sci.}, \textbf{99}, 136--148, 2014.

Hess, S.\,L.\,G., L. Lamy, and B. Bonfond, ISaAC, a Jupiter magnetic field model constrained by the auroral footprints of the Galilean satellites, in \textsl{Planetary Radio Emissions VIII}, edited by G. Fischer et al., Austrian Academy of Sciences Press, Vienna, this issue, 2017a.

Hess, S.\,L.\,G., P. Zarka, B. Cecconi, and L. Lamy, ExPRES: a tool to simulate planetary and exoplanetary radio emissions, \textsl{Astron. Astrophys.}, submitted, 2017b.

Higgins, C.\,A., Satellite control of Jovian 2--6~MHz radio emission using Voyager data, \textsl{J. Geophys. Res.}, \textbf{112}, A05213, 2007.

Hospodarsky, G.\,B., I.\,W. Christopher, J.\,D. Menietti, W.\,S. Kurth, D.\,A. Gurnett, T.\,F. Averkamp, J.\,B. Groene, and P. Zarka, Control of Jovian radio emissions by the Galilean moons as observed by Cassini and Galileo, in \textsl{Planetary Radio Emissions V}, edited by H.\,O. Rucker, M.\,L. Kaiser and Y. Leblanc, Austrian Academy of Sciences Press, Vienna, 155--164, 2001.

Kaiser, M.\,L., and J.\,K. Alexander, Periodicities in the Jovian decametric emission, \textsl{Astrophys. Lett.}, \textbf{14}, 55, 1973.

Kivelson, M.\,G., F. Bagenal, W.\,S. Kurth, F.\,M. Neubauer, C. Paranicas, and J. Saur, Magnetospheric interactions with satellites, in: Jupiter: the planet, satellites, and magnetosphere, edited by F. Bagenal,  W. McKinnon, T. Dowling, Cambridge University Press, pp. 513--536, 2004.

Lamy, L., P. Zarka, B. Cecconi, L. Klein, S. Masson, L. Denis, A. Coffre, and C. Viou, 1977--2017: 40 years of decametric observations of Jupiter and the Sun with the Nan\c cay Decameter Array, in \textsl{Planetary Radio Emissions VIII}, edited by G. Fischer et al., Austrian Academy of Sciences Press, Vienna, this issue, 2017.

Louis, C. K., L. Lamy, P. Zarka, B. Cecconi, S. Hess, and X. Bonnin, Detection of Jupiter decametric emissions controlled by Europa and Ganymede with Voyager/PRA and Cassini/RPWS, confirmed by Nan\c cay and Wind/Waves, in \textsl{Planetary Radio Emissions VIII}, edited by G. Fischer et al., Austrian Academy of Sciences Press, Vienna, this issue, 2017a.

Louis, C. K. , L. Lamy, P. Zarka, B. Cecconi, and S.\,L.\,G. Hess, Detection of Jupiter decametric emissions controlled by Europa and Ganymede with Voyager/PRA and Cassini/RPWS, \textsl{J. Geophys. Res.}, \textbf{in press, 2017b.}

Marques, M.\,S., P. Zarka, E. Echer, V. B. Ryabov, M. V. Alves, L. Denis, and A. Coffre, Statistical analysis of 26 years of observations of decametric radio emissions from Jupiter, \textsl{Astron. Astrophys.}, \textbf{604}, A17, 2017.

Menietti, J.\,D., D.\,A. Gurnett, W.\,S. Kurth, and J.\,B. Groene, Control of Jovian radio emission by Ganymede, \textsl{Geophys. Res. Lett.}, \textbf{25}, 4281, 1998.

Menietti, J.\,D., D.\,A. Gurnett, and I. Christopher, Control of Jovian radio emission by Callisto, \textsl{Geophys. Res. Lett.}, \textbf{28}, 3047, 2001.

St. Cyr, O.\,C., Jupiter's decameter and kilometer emissions: satellite effects and long term periodicities, PhD Thesis, University of Florida, Gainesville, USA, 1985.

Zarka, P., Auroral radio emissions at the outer planets: Observations and theories, \textsl{J. Geophys. Res.}, \textbf{103}, 20159--20194, 1998.

Zarka, P., Plasma interactions of exoplanets with their parent star and associated radio emissions, \textsl{Planet. Space Sci.}, \textbf{55}, 598--617, 2007.

Zarka, P., Star--Planet Interactions in the Radio Domain: Prospect for Their Detection, in "Handbook of Exoplanets" (Planets and their Stars: Interactions), J. Antonio Belmonte and H. Deeg Eds-in-chief, Springer, in press, 2017.

Zarka, P., M. Soares--Marques, C. Louis, V.\,B. Ryabov, L. Lamy, E. Echer, and B. Cecconi, Energetics of the Ganymede--Jupiter radio emission and consequence for radio detection of exoplanets, submitted, 2017.

\end{document}